\documentclass[conference]{IEEEtran}
\IEEEoverridecommandlockouts
\usepackage{cite}
\usepackage{amsmath,amssymb,amsfonts}
\usepackage{algorithmic}
\usepackage{graphicx}
\usepackage{textcomp}
\usepackage{xcolor}
\usepackage{hyperref}

\def\BibTeX{{\rm B\kern-.05em{\sc i\kern-.025em b}\kern-.08em
    T\kern-.1667em\lower.7ex\hbox{E}\kern-.125emX}}
\begin{document}

\bstctlcite{BSTcontrol}

\title{LiveData\\A Worldwide Data Mesh for Stratified Data}

\author{\IEEEauthorblockN{Simone Bocca}
\IEEEauthorblockA{\textit{Department of Information Engineering} \\ \textit{and Computer Science} \\
\textit{University of Trento}\\
Trento, Italy \\
simone.bocca@unitn.it}
\and
\IEEEauthorblockN{Amarsanaa Ganbold}
\IEEEauthorblockA{\textit{School of Information Technology} \\ \textit{and Electronics} \\
\textit{National University of Mongolia}\\
Ulaanbaatar, Mongolia \\
amarsanaag@num.edu.mn}
\and
\IEEEauthorblockN{Tsolmon Zundui}
\IEEEauthorblockA{\textit{School of Information Technology} \\ \textit{and Electronics} \\
\textit{National University of Mongolia}\\
Ulaanbaatar, Mongolia \\
tsolmonz@num.edu.mn}
}

\maketitle

\begin{abstract}

Data reuse is fundamental for reducing the data integration effort required to build data supporting new applications, especially in data scarcity contexts. However, data reuse requires to deal with data heterogeneity, which is always present in data coming from different sources. Such heterogeneity appears at different levels, like the language used by the data, the structure of the information it represents, and the data types and formats adopted by the datasets. Despite the valuable insights gained by reusing data across contexts, dealing with data heterogeneity is still a high price to pay. Additionally, data reuse is hampered by the lack of data distribution infrastructures supporting the production and distribution of quality and interoperable data. These issues affecting data reuse are amplified considering cross-country data reuse, where geographical and cultural differences are more pronounced. In this paper, we propose LiveData, a cross-country data distribution network handling high quality and diversity-aware data. LiveData is composed by different nodes having an architecture providing components for the generation and distribution of a new type of data, where heterogeneity is transformed into information diversity and considered as a feature, explicitly defined and used to satisfy the data users purposes. This paper presents the specification of the LiveData network, by defining the architecture and the type of data handled by its nodes. This specification is currently being used to implement a concrete use case for data reuse and integration between the University of Trento (Italy) and the National University of Mongolia.

\end{abstract}


\section{Introduction}



        Digital data is, nowadays, a fundamental resource for research and innovation development. The Big Data market grown fast in the last decade, but also Small Data became crucial for data-based applications, data analysis, as well as for AI applications, especially in contexts affected by data scarcity. In this scenario, data reuse can significantly reduce the cost of the inevitable data management process for getting valuable information from existing data. One of the key activities involved in such a process, in order to reap the benefits of data reuse, is the ability to integrate data from different data sources. More specifically, the aim of data integration is to unify and merge all the information carried by data coming from different sources by combining the semantic (knowledge representation) and syntactic (data value representation) aspects of the data in a purposeful form.

        
        Nevertheless, the reuse of data, and by consequence the data integration, directly involve dealing with the data heterogeneity, always present in data provided by existing data sources. Such heterogeneity occurs at different levels within the data to be reused. At syntactic level it appears in the file formats and in the data types adopted to represent specific information, while at semantic level it appears in the representation of real world entities structured by different properties according to the point of view of the data producer. This phenomenon significantly increases when the data reuse is considered worldwide, thus trying to reuse, and integrate, data from very different geographical and social contexts. For example, enabling data re-use data between Italy and Mongolia requires overcoming the problem of the different languages used to represent the data in the two countries, namely the language heterogeneity \cite{giunchiglia2017understanding, giunchiglia2018one}. At another level, for example, the data representation of university staff in Italy uses different data properties to the data representation of university staff in Mongolia, because of the differences in the organisation of the different university roles in the two countries, such as professors, researchers and students. This last example refers to what is called knowledge heterogeneity. Despite the difficulties in dealing with data heterogeneity, there are several examples of cross-country data reuse, that generated valuable analytical insights in different research fields, healthcare \cite{kiourtis2022electronic, KD-2020-Bella3}, environmental analysis \cite{belletti2020more}, digital innovation \cite{poess2014tpc}, and many others.
        
        
        In such a heterogeneous international context, data re-use is hampered not only by the cost of data integration to deal with such data heterogeneity, but also by the lack of high quality data distribution to enable effective cross-country data sharing. The ability to reuse data is directly dependent on the accessibility of such data, as well as on the ability to recognise whether or not data is suitable for a purpose-specific reuse. Several open data projects \cite{molinari2014big}, as well as data catalogues \footnote{\href{https://data.europa.eu/en}{European Open Data Portal}} have been created to address such data distribution issue. However, they are often focused on a specific field, context or country, and therefore they rarely (if ever) consider the possibility of sharing data between countries with very strong cultural differences (heterogeneity). Metadata definition plays a crucial role in such a cross-country data distribution. Often the metadata used to describe the data does not provide enough information about the data, which actually limits its potential reuse. Moreover, very rarely does the metadata take into account the linguistic diversity of the users who might be interested in reusing the data, thus limiting the metadata definition to the language of the data producer.


        In this paper we propose LiveData, namely a cross-country data distribution network handling high quality, and diversity-aware data. The fundamental idea behind the design of LiveData is to see the data heterogeneity not as a limitation, but as a feature of the data used to highlight the valuable information diversity it contains. "The world is extremely diverse and diversity is visibly manifested in language, data and knowledge" \cite{KD-2012-Giunchiglia}; with this sentence, F. Giunchiglia et al. want to communicate that the same real-world object can be described by many different words in different communities and in different languages, and can be represented by different property sets, with different property values. However, diversity allows for local maximisation of information representation by minimising the effort required to exploit it locally. Turning the data heterogeneity into information diversity, and considering it as a feature, the goal of each node that makes up the LiveData network is to transform existing low quality data by making its diversity explicit at language, knowledge, and data value level. This multi-layered, or stratified \cite{KD-2021-Giunchiglia-Stratified-DI}, approach to diversity representation increases the amount of valuable information in a single piece of data, as well as its reusability. In addition, the diversity-aware data produced are distributed in a worldwide data distribution network, structured according to the principles of data mesh architecture \cite{machado2022data, dehghani2022data}, in which the distribution and ownership of the data are controlled locally by the domain experts who created the data, having the necessary knowledge to fully express its relative information (and information diversity). The key idea behind the design of such a data distribution network is to preserve diversity also at the architectural level by giving each node full control over the data it creates and distributes. The principles of data interoperability and data reuse behind the definition of the LiveData network and the diversity-aware data, have been presented by F. Giunchiglia et al. in \cite{giunchiglia2022architecture}. Nevertheless, in this paper we propose to extend this previous contribution by providing a multi-domain data distribution model.



        The rest of the paper is organized as follows. Section 2 describes the diversity-aware data handled by the LiveData nodes. In section 3 we define the architecture of the LiveData network, by describing the architecture of each node, as well as the data distribution policy defined to share data between such nodes. Section 4 focuses on the related work in terms of existing data distribution networks and data catalogues. Section 5 concludes the paper by summarizing the key features of LiveData, and outlying the future work.

\section{LiveData contents}

    When we browse the web or collect data from existing databases, most of the time we get data where some of the information is implicit and therefore not directly accessible. For example, looking at a dataset containing information about university professors, we can see a set of values representing the attributes of different entities (professors), and we can also understand what the specific attributes are that these values refer to, like "first name", "last name", "courses taught", and others. However, we could not always understand which language(s) was used to represent the values (for example, if Italian citizens deal with a Mongolian dataset, or vice versa), especially considering the existence of languages more exploited then others for digital solutions \cite{helm2024diversity}. In addition, it may not be clear from the dataset what the meaning of certain attributes is. For example, it is difficult to understand whether the attribute "courses taught" refers to master's courses, bachelor's courses or both. More generally, it is not clear whether the entities and their attributes are part of an ontological model. Moreover, context-specific datasets often use local data standards to represent their information, making it difficult to clearly understand their data (for example, the values of the "courses taught" attribute may be represented by codes defined locally by the specific university). The example above shows how diversity can be expressed at different levels in a single dataset, and how the current common types of datasets are not able to express such diversity concretely.

    Each node of the LiveData network is designed to handle diversity-aware data, namely data where diversity has been made explicit by producing a dataset package, having different dataset types, one for each level at which diversity occurs. This type of data is produced through a specific data transformation process, implemented by following the iTelos methodology \cite{KD-2022-Bocca, giunchiglia2022popularity,giunchiglia2021itelos}. Such a process is exploited by each LiveData node to transform existing low quality data into diversity-aware data with an high level of interoperability and reuse. Applying the iTelos methodology, the process takes low quality input datasets and produces four different types of output datasets: 

    \begin{itemize}
        \item \textit{Standardised Datasets}: these are datasets that have been cleaned and formatted by adopting well-known standards. Such datasets have been cleaned by the iTelos process to remove any kind of possible noise, namely data values (but also attributes, or entire datasets) with additional meaningless information. In addition, the datasets are formatted to conform to well-known and widely used data standards. The standards adopted for these datasets are in line with the principles of $5\star$ Linked Open Data \footnote{\href{https://5stardata.info/en/}{Linked Data}} and FAIR data \footnote{\href{https://www.go-fair.org/fair-principles/}{FAIR data}}. The standardised datasets, managed by the LiveData nodes aim to be highly interoperable and reusable, while preserving their original information.\\
        
        \item \textit{Language Datasets}: the second type of dataset handled by the LiveData nodes aims to make explicit the diversity at the language level \cite{giunchiglia2018one, bocca2022interoperating}. The language datasets contain the definition of the concepts used to express the information carried by the other (types of) datasets. Such concepts are used to express entity and attribute names that can be found in the datasets, as well as Entity Types (ETypes) and ETypes properties that exist in the ontologies used to model the information in the dataset. The language datasets provide three key elements for each concept. (i) A concept identifier that uniquely identifies the concept, (ii) the concept word that represents the concept, and (iii) the description of the concept in one or more natural languages that is used to disambiguate the meaning of the concept. Specifically, the language datasets can be defined in different formats, such as CSV and XML, depending on the purpose of their use. The language datasets can be used to improve the information expressed by lexical databases \cite{giunchiglia2023representing}, by providing new concept definitions for different languages. \\ 

        \item \textit{Knowledge Datasets}: the knowledge dataset handled by the LiveData nodes aims to make explicit the diversity at the knowledge level. Such datasets are the ontologies produced by the iTelos process, which make explicit, and machine-readable, the ontological model of the information as it is implicitly represented in the datasets \cite{giunchiglia2023knowledge}. More in details, the knowledge datasets define how the ETypes are modelled by data properties and how they are related to each other by object properties. In addition, such ontologies define the functional level of each EType by expressing how each of them represents a specific function of a more general EType. For example, the representation of the EType "course taught" can better reflect the diversity of the specific context in which it has to be used, through an ontology modelling two more specific functional ETypes, like "master course" and "bachelor course". The LiveData knowledge datasets, defined as RDF-OWL files, are modelled reusing well-known standard reference ontologies, with the aim of enhancing the interoperability and the reuse of the newly created ontologies.\\    

        \item \textit{Graph-based Datasets}: the graph-based dataset \cite{angles2008survey} is the last type of datasets, by considering the generation order of the iTelos process. The key idea behind the generation of such datasets is to provide a type of data that contains all the types of datasets described above, thus making the different levels of diversity explicit in a single object. More specifically, graph-based datasets are Knowledge Graphs (KGs) \cite{K-2021-kejriwal}, which have a schema defined by LiveData knowledge resources, where the ETypes and their properties have been defined using the concepts of language datasets. The data layer of the KG, on the other hand, is defined using the LiveData standardised datasets. The merging of language, knowledge and standardised datasets produces a type of data in which diversity is concretely defined at different levels and in a machine-readable form. Furthermore, the interoperability of the type of datasets described above is directly reflected in the graph-based datasets, allowing the definition and distribution of more reusable KGs. The LiveData graph-based datasets are specifically defined as RDF Turtle (TTL) files.


    \end{itemize}

    It is important to note that LiveData creates and distributes content that can be both composed into a single object (the graph-based dataset) and decomposed into three different types of data (language, knowledge and standardised datasets). Such composition and decomposition allows the LiveData user to reuse the provided content as a whole or just one (or more) of its components, depending on their own specific purpose. This approach reduces the processing effort required to extract a single content's type from a single dataset object, as well as the effort required to build a KG from different datasets. In addition, such a compositional approach can also be used by considering content created from different input datasets, thus increasing the reuse of datasets across purposes.        
    To support the distribution of its content, each LiveData node uses the iTelos process to generate a specific set of metadata for each of the content types described above. A specific metadata schema is applied to each type of dataset to define the information to be provided by the metadata. The metadata plays a crucial role in the distribution of the data provided by the LiveData nodes. The information provided by the metadata allows LiveData users to find the content they are looking for, and to verify that such content is suitable for their own purposes. In other words, metadata increases the reusability of the data.

\section{LiveData architecture}

\begin{figure}[htbp]
        \centerline{\includegraphics [width=2.8in]{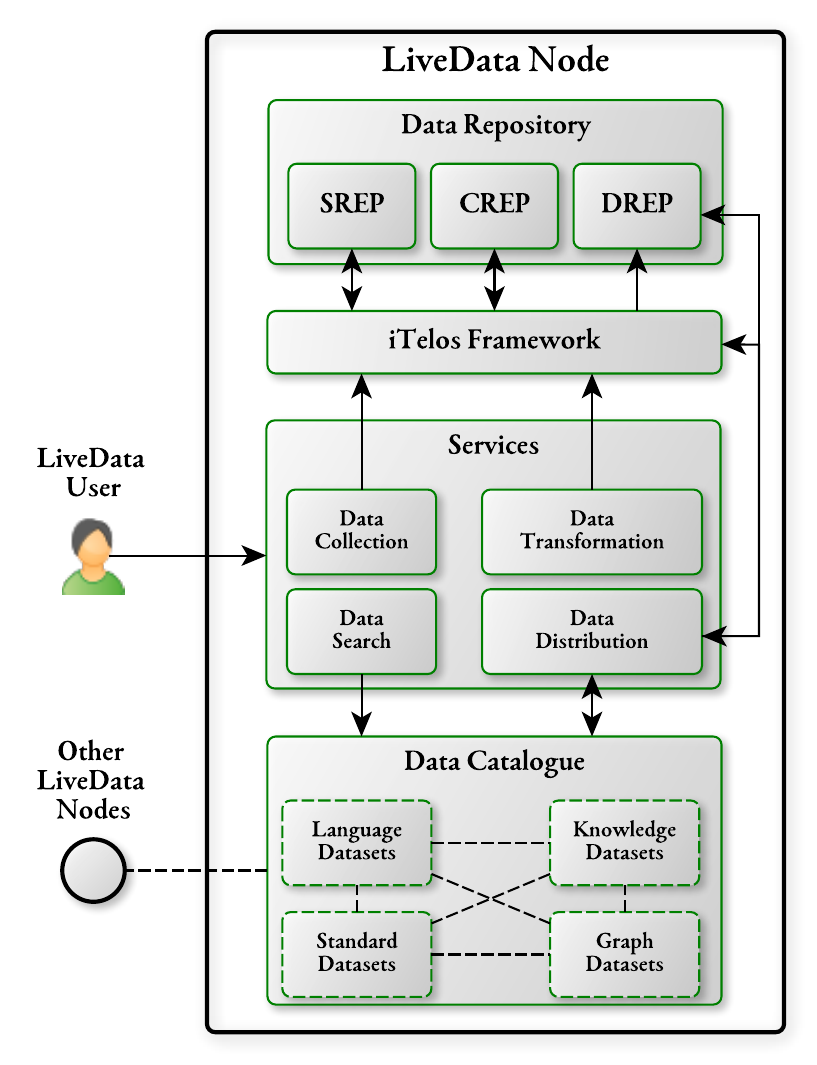}}
        \caption{LiveData Node Architecture}
        \label{fig:node}
    \end{figure}

    LiveData is a worldwide data distribution network sharing diversity-aware contents. To this end, the LiveData network architecture model has been designed as a Data Mesh \cite{dehghani2022data}. In their work, Machado et al. \cite{machado2022data} describe the four main principles of the data mesh model as (i) decentralising data management and ownership into the hands of domain representatives; (ii) defining an infrastructure that supports data as a product; (iii) defining quality and security data compliant with their requirements; and (iv) defining interoperable data. While the third and fourth data mesh principles are satisfied by the iTelos process executed by the LiveData nodes to generate diversity aware data, the first and second principles require the design, and implementation of a specific infrastructure. To this end, the LiveData data mesh architecture is composed of different nodes, defined for different geographical and cultural domains, which act as domain representatives, thus responsible for the maintenance and distribution of their local data. In addition, each node implements the concept of connectivity by defining links between the diversity-aware data thus highlighting such data as the main product provided. The architectural model of each LiveData node  depicted in Figure \ref{fig:node}. Each node is composed internally by four components described here below.

    The \textit{Data Repository} is the infrastructure's component used for the storage, and maintenance of the LiveData contents. The LiveData repository can be implemented by using several data management technologies ensuring data storage, as well as the data security and eventually the data anonymization (in case of sensible data). Nevertheless, the repository logical structure is uniquely defined and it has to be observed regardless of the technology used to implement it. Such a structure takes into account the four main types of content handled by LiveData, as well as the different phases of data management, from the data collection to the data distribution. For this reason the LiveData repository is internally structured in three main logical partitions, as depicted in Figure \ref{fig:node}.

    \begin{itemize}
        \item \textit{Source Repository (SREP)}: this repository partition contains the low quality data, collected from external sources that needs to be transformed into diversity-aware data. The data stored in the source repository has a lower quality and interoperability level. However, following the LiveData approach, SREP is again divided into three sub-sections for the storage of low quality datasets, external language datasets (not generated by the iTelos process) and external reference ontologies or data schemas. The SREP is the repository partition from which the iTelos data process receives its input.\\
        
        \item \textit{Core Repository (CREP)}: this repository partition instead contains the data that has been transformed by the iTelos data process. The CREP stores the LiveData content, so this partition is also internally organised to maintain standardised datasets, language datasets, knowledge datasets and graph-based datasets. The CREP is the repository partition used to store the output of the iTelos data process awaiting distribution.\\
        
        \item \textit{Distribution Repository (DREP)}: the last repository partition contains the diversity-aware data ready for distribution. When the LiveData administrator wants to distribute a data stored in CREP, a copy of that data is stored in DREP. In other words, DREP contains a subset of the data stored in CREP. In addition, for each dataset copied into DREP, a dedicated set of metadata is created and stored into the same repository partition. It is important to note that in Figure \ref{fig:node}, DREP is directly connected to the LiveData distribution service, which is responsible for publishing metadata to the LiveData catalogues.
    \end{itemize}

    The \textit{iTelos Framework} is a set of data management libraries and tools designed to implement the iTelos methodology. Thanks to this framework, any LiveData node can execute the iTelos data transformation process to produce diversity-aware data. The framework aims to support the different phases of the iTelos methodology, starting from the data collection until the generation of the metadata for the data distribution, passing through the generation of all the LiveData dataset types. For this reason, in Figure \ref{fig:node}, the iTelos framework component is directly connected to SREP in order to handle the collection and storage of the low quality data to be transformed, but also to CREP to store, and eventually reuse, the produced diversity-aware data. Furthermore, the framework interacts with DREP to store the data and its associated metadata to be then distributed via the LiveData catalogues. The iTelos framework provides the tools used to execute some of the requests of the LiveData services that interact directly with the LiveData users or other LiveData nodes. More specifically, as shown in Figure \ref{fig:node}, the framework handles the requests from the data collection and data transformation services.\\
    
    The \textit{Services} are the components of the LiveData node architecture that allows LiveData users to interact with the node, thus actually implementing the communication between the LiveData network and its users. The LiveData users can be both the administrators of the nodes, who use the LiveData services to collect, transform and distribute new data, and users who are not involved in the administration of the LiveData network. The latter category of users are interested in using the LiveData services to browse and download the diversity-aware data available through the LiveData catalogue. To support the needs of such users each LiveData node provides four services, as depicted in Figure \ref{fig:node}.

    \begin{itemize}
        \item \textit{Data collection}: the data collection service is mainly used by the LiveData node administrator to collect local low-quality data with the aim of transforming it into diversity-aware data and making it available in the LiveData network. Such a service, as shown in Figure \ref{fig:node}, is connected to the iTelos framework, which provides the tools to implement the collection of data to be stored in SREP.\\

        \item \textit{Data Transformation}: the data transformation service is subsequently used by the LiveData node administrator, to transform, the data collected in SREP, into diversity-aware data. The service offers to the user an access point for the usage of the transformation tools provided by the iTelos framework, to generate the standardised datasets, the language datasets, the knowledge datasets and the graph-based datasets, to be stored in CREP.\\
        
        \item \textit{Data Distribution}: the data distribution service is used by the LiveData node administrator to copy the data to be distributed into DREP and to generate the associated metadata. The service then sends the generated metadata to the LiveData catalogue for online publication. Moreover, the data distribution service is used by non-administrator users for the download of the data stored in DREP, identified and accessed through the data catalogue. The iTelos framework provides the tools to support the functionalities of such a service.\\
        
        \item \textit{Data Search}: the data search service is mainly used by non-administrator users to look for the diversity-aware data suitable for their purpose. Such a service links the users directly to the LiveData catalogue, allowing them to browse the data by exploiting the values of the metadata defined to describe it.
    \end{itemize}

    The \textit{Data catalogue} is the last component of the LiveData node architecture. It is responsible for publishing diversity-aware data to be distributed in the LiveData network. The data catalogue is a web portal designed to provide information at three different levels. At a more general level, the catalogue describes, through its landing page, the geographical and cultural domain for which it has been created. To this end, it provides a general description of all the data published in the catalogue, as well as a description of the organisation responsible for these data (the agent that acts as the LiveData node administrator). The second level of information provided by the catalogue, is given by the list of the available distributed contents. This information is provided by a second catalogue web page that list all the datasets (for each type of LiveData content). In this web page, the LiveData catalogue provides the search bar linked by the data search service, as well as the possibility to filter the list of available datasets by domain-specific categories (some examples of data categories are geographical data, health data, education data, and many others). Moreover, the catalogue at this level, allow users to filter the datasets according to the type of LiveData content, so that only language datasets, knowledge datasets, standardised datasets, graph datasets or a combination of these dataset types can be displayed in the dataset list. The third, and more specific, level of information provided by the LiveData catalogue, is focused on the individual dataset. The catalogue provides a web page where a single dataset is described by its own set of metadata. The metadata are visualised on this web page to provide a detailed description of the dataset, thus helping the users to understand if the dataset is what they are looking for. The metadata visualized on this page are also the parameters taken into account by the search service, provided on the previous catalogue's web page. Moreover, in the last web page, the catalogue also provides a link for the dataset download. The data download can be implemented in different ways depending on the data distribution policy defined by the data owner (for example one LiveData node can implement this by submitting a download request to the data owner, while another node may implement this by an automatic download procedure). The metadata published in the LiveData catalogue (depicted in Figure \ref{fig:node} by the dashed boxes inside the data catalogue component) that defines the diversity-aware data is not only descriptive. In fact, specific metadata values are defined to link the catalogue web pages describing datasets generated by the same execution of the iTelos process. To better understand such a catalogue feature, we define the iTelos transformation of a given low-quality dataset, named X, into the diversity-aware data package, composed of the four LiveData content types, where S is the standardised dataset, L is the language dataset, K is the knowledge dataset and G is the composed graph-based dataset. The metadata set published on the catalogue web page for dataset G, contains the link to the catalogue web pages describing datasets L, K and S, thus linking the graph-based dataset to the LiveData contents used to build it. Following the same approach dataset K has specific metadata linking to the web page for dataset L, which has been used to define its ETypes and properties names. Such connections between the LiveData contents, represented by dashed lines in Figure \ref{fig:node}, implement at the catalogue level, the compositionality of the contents handled by LiveData. It is important to note how, the metadata defined for a dataset into the current LiveData node, can link to the dataset's web page published on the catalogue of another node, since the external dataset has been reused to generate a content of the current LiveData node (for example a knowledge dataset of one node can be reused to generate the graph-based dataset of other nodes). For this reason, in the Figure \ref{fig:node}, the LiveData catalogue is linked to other LiveData nodes by a dashed line, describing the metadata links between two or more nodes.



    
    
\section{Related work}


Previous work has addressed the issue of data distribution at different levels. At the transnational level, the European Data Portal\footnote{\href{https://data.europa.eu/en}{data.europa.en/en}} (EDP) is one of the most interesting examples. The EDP aims to disseminate data and link existing data catalogues and web portals across European countries. The EDP, uses the DCAT-AP\footnote{\href{https://www.w3.org/TR/vocab-dcat/}{www.w3.org/TR/vocab-dcat}} (Data Catalogue Vocabulary - Application Profile)  as its metadata standard. DCAT-AP is a specification based on the Data catalogue Vocabulary (DCAT) for describing European public sector datasets. Its purpose is to facilitate interoperability and enable a more efficient data exchange and publication across the European Union. On the metadata definition side, DCAT is an RDF vocabulary designed to facilitate interoperability between web-based data catalogues. The W3C recommends this standard and is widely adopted for data description. It allows for uniformly categorizing and cataloging of datasets, making it easier for data portals to share and integrate data from various sources. The DCAT-AP standard specifically tailors DCAT for use within the context of European data portals, ensuring consistency and compatibility across the data catalogues of EU member states. It includes a set of classes and properties to be used by public sector portals to describe datasets in a standardized way. This standardization facilitates the aggregation of datasets published in national, regional, or thematic catalogues, enabling users to find and access datasets more efficiently across Europe. While the portal may not offer direct data transformation tools, users typically leverage external tools and libraries for data transformation tasks. It also distributes two types of data (knowledge and standardized datasets). However, the EDP lacks a specific methodology for generating high quality and interoperable data. We can see this lack by observing how EDP data is generated according to local standards, often defined locally by European countries. This results in high costs for data re-use and integration when dealing with data provided by the EDP.

At national level instead, several countries governments, as well as public and private organizations, developed their own solution to distribute data. For example, the open data portal of the National University of Mongolia \footnote{\href{https://data.num.edu.mn/}{data.num.edu.mn}} not only distributes tabular datasets but also publishes a knowledge graph using semantic data integration methods \cite{amarsaag2022mmt,amarsaag2022mmt1}, making the data more flexibly usable. Another example of local data distribution is the Open Data Trentino (ODT) \cite{bedini2014open} portal \footnote{\href{http://dati.trentino.it/}{dati.trentino.it}}, publishing data about the Trento autonomous province of Italy. The portal is one of the first national data catalogues for the dissemination of open data and thus represents an important step towards supporting data reuse and integration. However, even these national projects do not emphasise the diversity of the data. In addition, at the infrastructure level, these projects have a centralised approach to data management and distribution, which requires more effort to scale up to include new heterogeneous data domains.

From an architectural point of view, different data network projects (some example are the work described in \cite{roehrs2017omniphr} and \cite{ahlgren2016internet}, but many others can be found) have been developed for data sharing and distribution, going beyond the European Data Portal to encompass a wide range of initiatives, platforms and technologies designed to facilitate data access, sharing and analysis across different sectors and domains. Current data networks focus more on data standardization, accessibility, and interoperability, with a centralized or federated approach to data distribution. They have also made significant strides in making data accessible and interoperable on a large scale. Nevertheless, such kind of solutions, do not relay on the same methodology for data creation, as well as they do not share data access and data distribution polices, thus resulting into a domain-specific isolated environments, where a huge cost have to be paid for the integration and reuse of data.


\section{Conclusion}

In this paper we presented LiveData a worldwide data mesh for the generation and distribution of diversity-aware data. The key idea behind the design of the LiveData architecture and the data it handles, is that the data heterogeneity should not be seen as a constraint that generates effort and cost, but instead should be transformed into data diversity that enhances the power of the information carried by domain-specific data. The data handled by LiveData is a new type of data, created from existing low-quality data, and shaped as a data package composed of four different datasets where the diversity of information is concretely represented, and therefore exploitable, at several different levels. LiveData is designed as a data network composed of different nodes with a specific architecture composed of different components. These components are: a data repository for storing all the data handled by each node; a framework composed of various tools implementing the iTelos methodology supporting the generation and distribution of diversity-aware data; a set of services allowing LiveData users to interact with the LiveData node; and a data catalogue for the distribution of diversity-aware data. LiveData provides a novel solution for reducing the cost of data reuse by considering a new definition of data and a new architecture for its distribution. The implementation of the first LiveData network use case is currently underway. The first two nodes have been already defined for the University of Trento (Italy) and the National University of Mongolia, which are accessible through their catalogues, called LiveData UNITN \footnote{\href{https://datascientiafoundation.github.io/LiveDataUNITN/}{LiveData UNITN}} and LiveData NUM \footnote{\href{https://datascientiafoundation.github.io/LiveDataNUM/}{LiveData NUM}}, respectively. These first two nodes have been defined to produce and distribute diversity-aware data in the university domain. In particular, the graph-based datasets produced by the two nodes can be composed to highlight different types of interactions between the two universities, such as co-authorship of papers, collaboration on research projects, inter-faculty educational courses, and others. The aim for the future is to increase the amount of diversity-aware data available in these first two LiveData nodes, and to create new nodes to enhance the sharing of data and knowledge across the LiveData network.


\section*{Acknowledgements}

The results described in this paper have been achieved also based to the research previously carried out by PhD students in the Knowdive research group of the Department of Engineering and Computer Science at the University of Trento (Italy). In particular, important contributions have been made by the forthcoming PhD thesis of Mayukh Bagchi and Xiaoyue Li, who are studying a knowledge representation methodology for the generation of interoperable ontologies and a method for the representation of geographical and social contexts through geographical data, respectively. Furthermore, PhD student Alessio Zamboni made a strong contribution with his research on the definition of a toolkit to support the iTelos methodology, which will be described in his forthcoming PhD thesis. A special thanks go to Khuyagbaatar Batsuren from the National University of Mongolia for his research on language diversity. The work described here is part of the long term initiative \textit{DataScientia} \footnote{\href{https://datascientia.disi.unitn.it/}{https://datascientia.disi.unitn.it/}}. DataScientia's main aim and purpose is to contribute to and influence the future technological evolution of Data-Centric Human-Aware Artificial Intelligence (AI) as it applies to everybody and everything, and in particular to people and society. We also thank the results and datasets published through the API of the National University of Mongolia-funded Digital NUM pilot project (P2022-4222).

\bibliographystyle{IEEEtran}

\end{document}